\documentclass[prd, amsfonts, onecolumn, nofootinbib, showpacs]{revtex4}
\usepackage{graphicx, epsfig}
\usepackage{color}
\usepackage{amsmath}
\usepackage{amssymb}
\usepackage{physics}
\newcommand{\be}{\begin{equation}}
\newcommand{\ee}{\end{equation}}
\newcommand{\bea}{\begin{eqnarray}}
\newcommand{\eea}{\end{eqnarray}}

\newcommand{\gapp}{\mathrel{\raise.3ex\hbox{$>$}\mkern-14mu
\lower0.6ex\hbox{$\sim$}}}
\newcommand{\lapp}{\mathrel{\raise.3ex\hbox{$<$}\mkern-14mu
\lower0.6ex\hbox{$\sim$}}}
\def\bbox{{\,\lower0.9pt\vbox{\hrule \hbox{\vrule height 0.2 cm
\hskip 0.2 cm \vrule  height 0.2 cm}\hrule}\,}}

\begin{document}
\title{Testing ER=EPR}
\author{De-Chang Dai$^{1,4}$\footnote{communicating author: De-Chang Dai,\\ email: diedachung@gmail.com\label{fnlabel}}, Djordje Minic$^2$, Dejan Stojkovic$^3$, Changbo Fu$^{5}$}
\affiliation{$^1$ Center for Gravity and Cosmology, School of Physics Science and Technology, Yangzhou University, 180 Siwangting Road, Yangzhou City, Jiangsu Province, P.R. China 225002 }
\affiliation{ $^2$ Department of Physics, Virginia Tech, Blacksburg, VA 24061, U.S.A. }
\affiliation{ $^3$ HEPCOS, Department of Physics, SUNY at Buffalo, Buffalo, NY 14260-1500, U.S.A.}
\affiliation{ $^4$ CERCA/Department of Physics/ISO, Case Western Reserve University, Cleveland OH 44106-7079}
\affiliation{ $^5$ Institute of Modern Physics, Fudan University, Shanghai, 200433, P.R. China}

\begin{abstract}
\widetext
We discuss a few tests of the ER=EPR proposal. We consider certain conceptual issues as well as explicit physical examples that could
be experimentally realized. In particular, we discuss the role of the Bell bounds, the large N limit, as well as the consistency of certain
theoretical assumptions underlying the ER=EPR proposal. As explicit tests of the ER=EPR proposal we consider limits coming
from the entropy-energy relation and certain limits coming from measurements of the speed of light as well as measurements of
effective weights of entangled states. We also discuss various caveats of such experimental tests of the ER=EPR proposal.
\end{abstract}


\maketitle

\section{Introduction}

In a very interesting, intriguing and inspiring attempt to connect the geometry of space-time and fundamental properties of quantum mechanics Maldacena and Susskind  proposed the so-called ER=EPR conjecture \cite{Maldacena:2013xja}. It is basically an assertion that two entangled \cite{epr}, \cite{bell}
quantum particles are also connected by a non-traversable wormhole \cite{wormhole}.  If this conjecture turns out to be true (see also \cite{holland}), it would indeed revolutionize the way we are thinking about unifying quantum mechanics with gravity 
(see, for example, \cite{VanRaamsdonk:2010pw}). It is therefore very important to verify its self-consistency and find a way to test it in the relevant experimental environment. 

The ER=EPR conjecture has some immediate implications, and even concrete qualitative consequences and applications  (see, for example, \cite{Brown:2019hmk}). So far no quantitative predictions have been checked, even though there are some count-examples\cite{Chen:2016nvj} and some proposals\cite{Chen:2016xqz,Chen:2017cgw}. This is partially because there are still no concrete realizations of this conjecture in terms of realistic physical systems. Nevertheless, we argue here that even the basic premises of the ER=EPR conjecture allow us to to put very strong constraints that must be overcome in any realistic realizations of this idea. 
In light of these constraints we also discuss various caveats regarding the proposed experimental tests of the ER=EPR proposal.

\section{The ER=EPR setup}

In this section we summarize the essential assumptions of the ER=EPR setup. 
The ER=EPR proposal was essentially motivated by the picture \cite{Maldacena:2001kr} of eternal black holes in the context of the AdS/CFT dictionary \cite{Maldacena:1997re}, including
the following two key assumptions explicitly spelled out in \cite{Maldacena:2013xja}

\begin{itemize}

    \item[{\bf 1.}] {\bf Two entangled particles are connected by a wormhole}
    
(We quote from page $2$ in \cite{Maldacena:2013xja}:) 

{\it It  is  very tempting  to  think  that any EPR  correlated  system  is  connected  by  some  sort  of  ER bridge, although in general the bridge may be a highly quantum object that is yet to be independently defined.  Indeed, we speculate that even the simple singlet state of two spins is connected by a (very quantum) bridge of this type.} 

    \item[{\bf 2.}]{\bf If two particles are not maximally entangled, then a wormhole that connects them has two disconnected horizons}

(We quote from page $12$ in \cite{Maldacena:2013xja}:)

{\it 2.6    Bridges for less than maximal entanglement

In the Penrose diagrams we have discussed the Left and Right horizons touch each other.It is also possible to have configurations where they do not touch each other.  A simple way to generate them is to start from two eternal black holes and add some matter to each side.  These configurations can also be prepared by considering Euclidean evolution with a time dependent Hamiltonian, see [22] for some explicit solutions.  The Penrose diagram of such configurations is given in figure 7} \cite{Maldacena:2013xja}.

(We quote from the page $13$ in \cite{Maldacena:2013xja}:)

{\it Figure 7:   Penrose diagram of a configuration obtained by analytic continuation of a time reflection symmetric, but time dependent, Euclidean solution.  The two horizons do not touch.  The entanglement, computed by the Ryu-Takayanagi prescription [23], is given by the area of a minimal surface with less area than the horizons. The area of the horizons grows when we go from the bifurcation point to the future.}

\end{itemize}

\section{Consistency of the key assumptions of the ER=EPR proposal}

A pair creation generates an entangled pair of particles, which according to  EPR=ER proposal generates a wormhole. It is then instructive to ask what happens to the wormhole when two particles that carry wormholes with them annihilate.

\begin{figure}
\includegraphics[width=10cm]{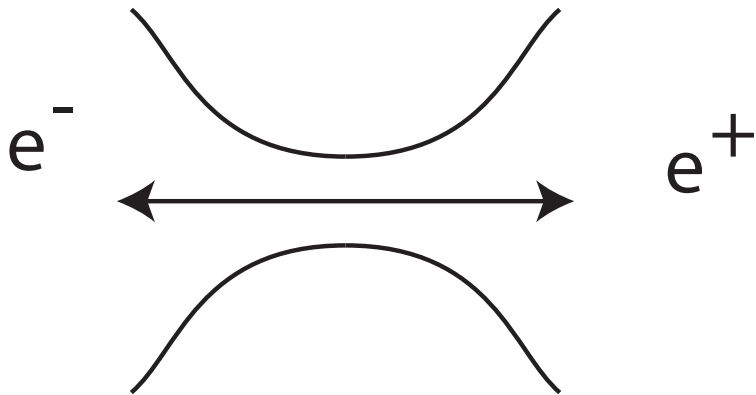}
\caption{An entangled $e^+e^-$ pair is created. There is an ER bridge connecting the particles.  
}
\label{entangle}
\end{figure}

\begin{figure}
\includegraphics[width=10cm]{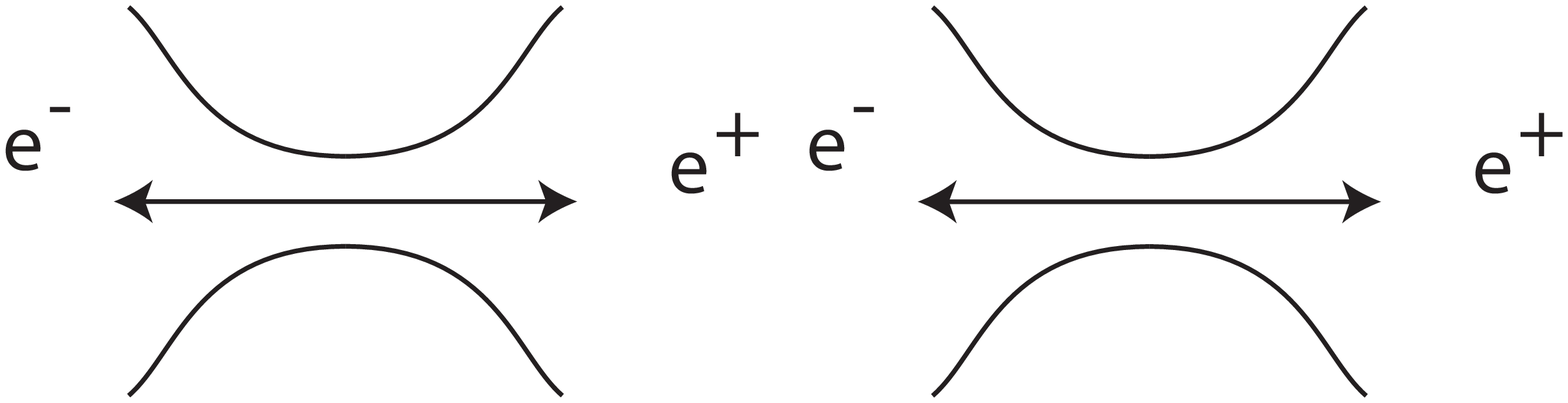}
\caption{Two entangled $e^+e^-$ pairs are created. There are now two ER bridges present in the space.
}
\label{entangle2}
\end{figure}

\begin{figure}
\includegraphics[width=10cm]{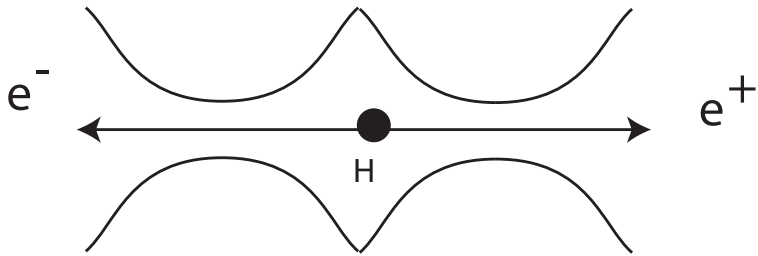}
\caption{ A positron from one pair and an electron from the other interact in the middle and create a Higgs particle. The electron and positron at the two ends must be almost maximally entangled. However, the two initial bridges are connected into one continuous configuration with the Higgs in the middle, so the two horizons at the end do not touch. This is not consistent with the assumption $2$, which implies that the horizons must touch in a maximally entangled state.      
}
\label{entangle3}
\end{figure}

We start with an observation that the two key assumptions 1 and 2 in the ER=EPR set-up are, apparently, not always consistent with each other. Consider an $e^+ e^-$ pair, created in a maximally entangled state (Fig.~\ref{entangle}).  The particles are connected by an ER bridge. Their spin state is 
\begin{equation}\label{epem}
\sim \ket{\uparrow_1 \downarrow_2}+\ket{\downarrow_1 \uparrow_2} .
\end{equation}
Suppose now that two such $e^+e^-$ pairs are created, each of them in a maximally entangled state described by  Eq.~(\ref{epem}), as in Fig.~\ref{entangle2}. There are now two ER bridges present in the space. The wavefunction describing this configuration is
\begin{equation}
\sim (\ket{\uparrow_1 \downarrow_2}+\ket{\downarrow_1 \uparrow_2} )(\ket{\uparrow_3 \downarrow_4}+\ket{\downarrow_3 \uparrow_4}) .
\end{equation}
Now, consider the following interaction 
\begin{equation}
e^++e^- \rightarrow H ,
\end{equation}
where $H$ is the Standard Model Higgs boson or any other scalar field. In the context of our configuration, the particle $2$ (a positron) may interact with the particle $3$ (an electron) and create a Higgs in the middle. At the tree level, particles $2$ and $3$ must have opposite spins. The same spin state is suppressed and, therefore, the new state is approximately

\begin{equation}
\sim \ket{\uparrow_1 \downarrow_4 H}+\ket{\downarrow_1 \uparrow_4 H}  .
\end{equation}
This implies that the particle $1$ (an electron) and particle $4$ (a positron)  are almost in the maximally entangled state. The wormhole bridge is not disconnected in this process - the two bridges just merge into one continuous configuration with the Higgs particle in the middle (Fig.~\ref{entangle3}). Now this new wormhole has two disconnected horizons at its ends, which according to the assumption $2$ in the previous section implies that the final $e^+e^-$ pair is not in the maximally entangled state. This is apparently an inconsistent situation, which perhaps requires modifying or sharpening the ER=EPR proposal. For example, it may happen that even maximally entangled states are connected with an elongated wormhole in some cases \cite{Shenker:2013yza}.

\section{Bell's bounds, ER=EPR, and the large N limit}

Another conceptual comment concerns the fundamental puzzle of (apparently)
mismatched Bell bounds \cite{bell} in the ER=EPR proposal.
The quantum Bell bound of an EPR entangled state is of order of magnitude larger than the naive classical
Bell bound on the ER side of the correspondence. Of course, one might invoke 
the AdS/CFT correspondence and, in particular, the large N limit, in which these two bounds might be
comparable for the correct gauge invariant observables, presumably because of large N factorization of the relevant correlators.
This would suggest the need for an AdS/CFT like
correspondence and the appropriate large N limit that would make the ER=EPR proposal operationally viable.
(Another possibility is that, generically, the wormhole involved is strongly quantum, and thus both sides of the ER=EPR relation would be saturating the quantum Bell bound.)

Let us now look at the large N limit \cite{Terashima:2019wed} in a bit more technical detail by following \cite{Maldacena:2001kr}. In the context of AdS/CFT the wavefunction for 
an EPR entangled pair can be written as (this is equation (2.10) in \cite{Maldacena:2001kr})

\begin{equation}
\ket{\Psi}=\frac{1}{\sqrt{Z(\beta ,\mu )}} \sum_n e^{-\frac{\beta E_n}{2}-\frac{\beta \mu l_n}{2}} \ket{E_n, l_n}_1 \ket{E_n,l_n}_2 ,
\end{equation}
or, equivalently, as equation (2.4) in \cite{Maldacena:2013xja}
\begin{equation}
\label{bhpair}
\ket{\Psi (t)}= \sum_n e^{-\frac{\beta E_n}{2}} e^{-2i E_n t}  \ket{\bar{n},n} ,
\end{equation}
(where $\bar{n}$ is CPT conjugate of state $n$). According to the ER=EPR proposal, $\ket{\Psi}$'s entanglement entropy is the same as the Bekenstein-Hawking entropy (We quote from the page $5$ in \cite{Maldacena:2013xja}:)

{\it The second interpretation of the eternal black hole is that it represents two black holes in disconnected spaces with a common time [7, 8, 9, 10].  We will refer to the disconnected spaces as sheets.  
The degrees of freedom of the two sheets do not interact but the black holes are highly entangled with an entanglement entropy equal to the Bekenstein-Hawking entropy of either black hole.  We say that these black holes are maximally entangled. }

The question is: can we calculate the Bell inequality for this state? In principle, one would have to find 3 or 4 measurable physical quantities, for example spin. 
This depends on what form of Bell inequality one is using. For example, 3 parameters are needed for the original Bell inequality, and 4 for its CHSH version \cite{bell}. 
Second, in the example  discussed in the previous section the particles that comprise an entangled pair are stable, i.e. $e^+$ and $e^-$.  However, if we consider unstable particles, for example 
\begin{equation}
\ket{W^+}\ket{W^-} \rightarrow \ket{\mu^+ \nu_\mu} \ket{e^- \bar{\nu }_e} ,
\end{equation} 
we need to be carefull when using equation (6).
If the left hand side of the above process is maximally entangled, then the product of this particular decay on the right hand side
should be maximally entangled, but that situation cannot be described by equation (6) because that equation assumes CPT conjugation,
and the right hand side of the above process violates that assumption. 
Therefore, it would be necessary to generalize equation (6) to include non-CPT states.

\section{Testing ER=EPR: limits from entropy-energy relation}

A possible quantitative test of the ER=EPR conjecture might come from the assertion that  the 
 entanglement entropy is given by the area of a minimal surface of the wormhole connecting two black holes in ER bridge.   Entanglement entropy of maximally entangled particles is equal to the Bekenstein-Hawking entropy of either black hole. 

\begin{itemize}
\item  (We quote from the page $20$ in \cite{Maldacena:2013xja}:)   

{\it The cut through the Einstein-Rosen bridge defines a two dimensional surface whose area should not be smaller than the entanglement entropy.  Based on [23] we expect that the smallest area of such a cut is, in fact, the entanglement entropy. }
\end{itemize}

\subsection{Entropy vs. Energy}
The Bekenstein-Hawking entropy is 
\begin{equation}\label{en}
S_{BH} = \frac{k_B c^3 A}{4G_N \hbar},
\end{equation}
where $A$ is the area of the cut through the Einstein-Rosen bridge. According to  \cite{Maldacena:2013xja}, this entropy must be lager than the entanglement entropy between the electron and positron in a pair, $ \ln (2) k_B$. The Schwarzschild radius associated with the mass $M$ is 
\begin{equation}
R_g=\frac{2G_N M}{c^2} ,
\end{equation} 
which gives the area of the cut as $A=4\pi R_g^2$. From Eq.~(\ref{en}), we obtain the ADM mass of the wormhole 
\begin{equation}\label{M}
M=\sqrt{\frac{ \hbar c\ln (2)}{4\pi G_N}}=5.11\times 10^{-9}kg = 2.87\times 10^{27}eV .
\end{equation}
Note that this value represents a minimal mass of the wormhole, since it does not account for its extended nature.  

This value is much larger than a typical elementary  particle pair's mass, e.g. $\sim MeV$ for an $e^+e^-$ pair. It seems highly implausible that such a wormhole can be created as a consequence of entanglement. In fact, we can immediately check that such a large wormhole energy grossly violates the existing data. For example, we can compare the energy of two free particles  with their  energy in an entangled state. The measurements for positronium are readily available.  In its ground state,  positronium is highly entangled. At the end of its lifetime, it decays to two entangled photons. When the photons are detected, they disentangle immediately. The excess entanglement energy (if present) should be released via gravitational and other channels when disentanglement happens (and perhaps used to establish new entanglement). The best current measurement of the energy of two photons coming  from positronium annihilation is $2m_0c^2 -2.4keV<E_1+E_2<2m_0c^2 +2.4keV$\cite{1977PhRvL..38..241L,1996PhRvL..77.2097A}. This is $24$ orders of magnitude smaller than the wormhole's effective mass in Eq.~(\ref{M}). Though the actual nature and mechanism of generating a wormhole in the ER=EPR conjecture is unknown, this enormous discrepancy in orders of magnitude severely complicates any concrete realization, and thus, it puts a very strong constraint on the proposed ER=EPR relation. 

The caveat here is that the relevant wormhole might be
a very quantum object, in which case one should be careful from drawing a strong conclusion from this example.

\subsection{Limits coming from the speed of light measurements}

An entangled particle carries a wormhole with it while it propagates through space. Then the mass/energy of a system wormhole+particle can be very different from the mass/energy of the host particle. If that particle is a photon, a wormhole that is dragged along would introduce an effective mass to the photon (due to interactions) and reduce its propagation speed. 
Therefore entangled photons must be slower than their disentangled cousins. The current experimental uncertainty of the speed of light according to \cite{1972PhRvL..29.1346E} is
\begin{equation} \label{c}
 \frac{\Delta c}{c}<3.5\times 10^{-9} .
 \end{equation}
 If an entangled photon's energy is $E$, we can reasonably assume that the energy to generate a wormhole must be lower than 
 $E$. Assuming that the effect of the wormhole must fit within the experimental speed of light uncertainty, the constraint in Eq.~(\ref{c}) can be translated into the constraint on the mass of the wormhole. Including the special relativistic effects, the rest mass of a wormhole must be 
\begin{equation} 
M <E\sqrt{1-\Big(\frac{c-\Delta c}{c}\Big)^2}\approx E\sqrt{2\frac{\Delta c}{c}} \approx 4\times 10^{-5} E .
\end{equation}
 
If, for example, an entangled photon is created by an electron transitioning from one to another atomic state, its energy in general is of the order of $eV$. This value puts a very stringent  bound on the energy of the wormhole, $<10^{-5}eV$. This is $32$ orders of magnitude less than what is expected from the entanglement entropy-energy relation we derived in Eq.~(\ref{M}). 

We can avoid this bound by assuming that a wormhole does not follow its host particle. However, that would also imply that the EPR-like entanglement is not addressed by introducing such a wormhole.  

\subsection{The weight of an entangled state}

If an entangled pair of particles actually hosts a wormhole of mass $M$, then the gravitational acceleration around this pair should be of the order of  
\begin{equation}
a\approx\frac{G M}{r^2}. 
\end{equation}
This means that the force between the Earth and this entangled pair is of the order 
\begin{equation}
F\approx\frac{G M M_E}{r^2},
\end{equation}
where $M_E$ is the Earth's mass. For $M= 4.06\times 10^{27}$eV from  Eq.~(\ref{M}), this pair's weight is $\approx 7.23\times 10^{-9}kg$ near the Earth's surface. This value is close to the test mass $1.4\, \mu$g used in the tests of Newton's force
at short distance \cite{Chiaverini:2002cb}. If we replace the gold rectangular prism in this experimental setup with an entangled pair, then we should be able to directly  test gravitational force generated by the wormhole. This experiment could be performed in the very near future. 

(Note that one might argue that the use of the above classical formula for the gravitational acceleration is not appropriate, and that this situation corresponds to the presence of a highly quantum wormhole, in which case one would have to use a
highly non-trivial quantum analog of the above classical formula, which, at present, is not known. This is one of the caveats listed in the next section. Nevertheless, the use of the classical formula for the gravitational acceleration does give us a feel for the numbers involved in this experimental set-up, and it does sharpen our physical understanding of the ER=EPR proposal.)

Alternatively, if we have electrons in an ion in the entangled state, for example $Li^+$, we can directly test its mass with a mass-spectrometer. One $Li^+$ ion is enough, because if two electrons are in the ground state, their spins must be entangled.The current mass-spectrometer precision reaches $\frac{\Delta m}{m}\sim 10^{-6}$ \cite{Marshall}. For a $Li^+$, this translates into $\Delta m = 5\times 10^{-27}kg$ or $3keV$.  This is, again, many orders of magnitude smaller than the estimated wormhole  energy needed for ER=EPR to work. This is already an underestimation, given a specific method the mass resolution may achieve $10^{-9}$ to $10^{-12}$\cite{Brodeur,Myers}. Thus, a much stringent constraint could be applied to this situation.

Finally, we can collect a large amount of entangled particle pairs. For example,
 two electrons in the $^4He$ ground state must be an entangled spin up and spin down pair. If they are connected by a wormhole, then their weight will significantly increase.   $^4He$ ground state atoms are, in principle, very easy to collect. For example, $10^9$ entangled $^4He$ electrons will weigh $7.23$ kg on Earth. This effect is virtually impossible to be missed in condensed matter experiments. 
Therefore, once again, we end up with a very strong constraint on the ER=EPR set-up, which leads us to
a summary of various caveats.

\section{Caveats: AdS/CFT and ER=EPR}

How are we to interpret the above strong constraints of the ER=EPR proposal?
In this section we list some obvious caveats to our discussion.
The ER=EPR proposal was really motivated by the picture of eternal black holes \cite{Maldacena:2001kr} in the context of the AdS/CFT dictionary \cite{Maldacena:1997re} .
In that context (and, in particular, the relation between eternal black holes in an asymptotically $AdS_3$ and the entangled boundary $CFT$s) it is intuitively very 
natural to expect that an entangled EPR-state of two boundary CFTs is represented geometrically by the Penrose diagram of an eternal black hole, which by definition, includes
the ER bridge. Thus, from this point of view the ER=EPR proposal should naturally follow from the AdS/CFT correspondence.
In all of the experimental tests presented in this letter one has physical realizations that are highly non-trivial from the point of view of the AdS/CFT correspondence.
Indeed, in the above proposed experiments the wormhole configuration must be a highly non-trivial quantum object, in order for the ER=EPR correspondence to make sense.
This is the main caveat to our discussion of the ER=EPR proposal. 

Another, more concrete way to evade these strong constraints, is to postulate that a fixed background geometry where our quantum particles are propagating might not be the same spacetime where the wormhole exists. For example, this may be realized in a variant of the so-called brane world models \cite{ArkaniHamed:1998rs,Randall:1999vf,Starkman:2000dy,Starkman:2001xu,Giudice:2016yja}. Such a setup would be more in the spirit of the GR=QM proposal 
\cite{Susskind:2017ney} (a far reaching extension of the ER=EPR correspondence) where quantum entanglement directly corresponds to classical geometry.   

In our discussion, we took the ER=EPR proposal at face value, which might be an oversimplification. Nevertheless, we think that the above experimental set-ups will help sharpen our understanding of this fascinating conjecture.

At the moment the ER=EPR proposal really makes sense in the context 
of AdS/CFT, as a reformulation of the description of eternal black holes in the 3 dimensional AdS bulk in terms of two entangled 2 dimensional CFTs. One might generalize this picture in any number of dimensions and outside of the AdS background as suggested by the ER=EPR proposal.
As we have indicated in previous sections, the latter seems to be hard to implement empirically. However, we could examine ER=EPR in a higher dimensional AdS example. 

Take the example of Wilson loop observables in 4d CFT as discussed in the classic papers \cite{WilsonMaldacena}. 
In that case the large N factorization property seems to be crucial in order to make sense of the ER=EPR proposal, and in particular the matching of the Bell bounds on both sides of the correspondence (as already alluded to previously). How about the large N corrections to this leading result? According to the ER=EPR proposal the quantum entanglement of heavy quarks degrees of freedom (at finite temperature) should correspond to the appearance of a ``wormhole" correction in the bulk.
The question is: can this work in detail?

Note that the entanglement of quantum degrees of freedom should be studied in the boundary CFT at finite temperature, because that corresponds to the presence of a black hole in the bulk AdS space. The cleanest higher dimensional case would involve quantum degrees of freedom in thermal N=4 SYM in 4d  and the black hole background in $AdS_5$.
This case comes closest to the real world of finite temperature QCD \cite{Son} and as such it could be used for a precise characterization of a ``quantum wormhole" that seems to be necessary in order to apply the ER=EPR empirically.
For example, entanglement entropy has been studied in this context in \cite{Lewkowycz:2013laa}.
In principle, one could study entanglement of a quark-anti-quark pair in the context of N=4 SYM theory using
the tools of the AdS/CFT correspondence, and then reformulate the result in terms of a generically non-trivial
``quantum wormhole'' configuration in the AdS bulk. This seems to realize the spirit of the ER=EPR proposal 
in a physical situation that is very close to the real world.
(And there is evidence for this in the existent literature \cite{Jensen:2013ora}.)
However, a detailed quantitative 
discussion of such a physical situation goes well beyond the scope of our present paper.


\section{Conclusion}

In this paper we have discussed a few tests of the ER=EPR proposal. We have concentrated our presentation on certain conceptual issues as well as explicit physical examples that could
be experimentally realized. In particular, we have discussed the role of the Bell bounds, the large N limit, as well as the consistency of certain
theoretical assumptions underlying the ER=EPR proposal. As explicit tests of the ER=EPR proposal, we have considered limits coming
from the entropy-energy relation and certain limits coming from the measurements of the speed of light and the measurement of
an effective weight of entangled states. The central take-home message of this letter is that 
the basic assumptions of the ER=EPR proposal allow us to to put very strong constraints that must be overcome in any realistic realizations of this very interesting idea. 
We note that the above phenomenological tests of the ER=EPR conjecture
could be, in principle, extended to the Planck scale, where the natural length scales in the
ER=EPR proposal (the gravitational and the Compton scale, respectively) are both of the order of the Planck scale, and then such a relation
could in principle pass various experimental constrains, but that situation would destroy most of its observational effectiveness.
Nevertheless,  the fundamental importance of non-locality (as exemplified by non-traversable wormholes featuring in the ER=EPR proposal) that is consistent with causality, is an essential message of 
quantum mechanics with fundamental implications for quantum field theory, string theory and quantum gravity (as pointed out and explored in \cite{Freidel:2013zga}), with observable effects that should be systematically investigated.

\begin{acknowledgments}
We thank V. Balasubramanian, R. G. Leigh, O. Parrikar, J. Simon for discussions. We also thank J. Maldacena and L. Susskind for email correspondence. Special thanks to O. Parrikar, J. Simon and J. Maldacena for comments on a draft of this
paper. D.C Dai is supported by the National Natural Science Foundation of China  (Grant No. 11775140). D. M. is supported in part by the US Department of Energy (under grant DE-SC0020262) and by the Julian Schwinger Foundation. D.S. is partially supported by the US National Science Foundation, under Grant No. PHY-1820738  and PHY-2014021.
\end{acknowledgments}


\end{document}